\documentclass[10pt,conference]{IEEEtran} 
\IEEEoverridecommandlockouts

\usepackage{cite}
\usepackage{multirow}
\usepackage{comment}
\usepackage{amsmath,amssymb,amsfonts}
\usepackage{algorithmic}
\usepackage{graphicx}
\usepackage{textcomp}
\usepackage{caption}
\usepackage{array}
\usepackage{tabularray}

\usepackage{ragged2e}
\usepackage[labelformat=simple]{subcaption}

\usepackage[table,xcdraw]{xcolor}
\def\BibTeX{{\rm B\kern-.05em{\sc i\kern-.025em b}\kern-.08em
    T\kern-.1667em\lower.7ex\hbox{E}\kern-.125emX}}
\begin{document}
\setlength{\textfloatsep}{0.3pt}

\title{Feasibility Study of Function Splits in RAN Architectures with LEO Satellites}%

\author{
 \IEEEauthorblockN{Siva Satya Sri Ganesh Seeram\IEEEauthorrefmark{1}, Luca Feltrin\IEEEauthorrefmark{2}, Mustafa Ozger\IEEEauthorrefmark{1}, 
Shuai Zhang\IEEEauthorrefmark{1}, Cicek Cavdar\IEEEauthorrefmark{1}}
\IEEEauthorblockA{\IEEEauthorrefmark{1} KTH Royal Institute of Technology, Sweden, Email: \{sssgse, ozger, shuai2, cavdar\}@kth.se \\
\IEEEauthorrefmark{2} Ericsson AB, Sweden, Email: luca.feltrin@ericsson.com
}
}

\maketitle

\begin{abstract}
This paper explores the evolution of Radio Access Network (RAN) architectures and their integration into Non-Terrestrial Networks (NTN) to address escalating mobile traffic demands. Focusing on Low Earth Orbit (LEO) satellites as key components of NTN, we examine the feasibility of RAN function splits (FSs) in terms of fronthaul (FH) latency, elevation angle, and bandwidth (BW) across LEO satellites and ground stations (GS), alongside evaluating performance of Conditional Handover (CHO) procedures under diverse scenarios. By assessing performance metrics such as handover duration, disconnection time, and control traffic volume, we provide insights on several aspects such as stringent constraints for Low Layer Splits (LLSs), leading to longer delays during mobility procedures and increased control traffic across the feeder link in comparison with the case when gNodeB is onboard satellite. Despite challenges, LLSs demonstrate minimal onboard satellite computational requirements, promising reduced power consumption and payload weight. These findings underscore the architectural possibilities and challenges within the telecommunications industry, paving the way for future advancements in NTN RAN design and operation.
\end{abstract}


\vspace{-2mm}

\section{Introduction}
\vspace{-1mm}

The evolution of Radio Access Network (RAN) architectures has been a key aspect in the advancements of wireless communication technologies. The escalating demand for mobile traffic within the conventional Distributed-RAN (D-RAN) paradigm catalyzes exploration and advancement toward novel RAN technologies~\cite{r1,r5}. Among these, the Centralized RAN (C-RAN) emerges as a proposed solution to address the intensifying requirements of mobile traffic. Nevertheless, it introduces substantial demands on the transport network, connecting the antenna site with the central site, concerning latency and bandwidth (BW)~\cite{r1}. Thus, a critical need exists to investigate solutions that reduce the capacity demands on this transport network. The concept of a Function Split (FS), refers to partitioning Baseband (BB) functions among different network elements to determine which functions remain at the antenna site and which ones are centralized. These FSs allow for greater flexibility and efficiency in network design and operation leading to novel architecture solutions~\cite{r1,r5,r4}.

The discussions so far on RAN architectures have primarily centered on terrestrial networks (TN), where users exhibit mobility while RAN deployment remains stationary. However, the integration of non-terrestrial networks (NTN) becomes imperative to achieve ubiquitous coverage. Components of NTN may encompass satellites, high-altitude platforms, low-altitude platforms, and other similar entities. Additionally, it is crucial to examine the feasibility of RAN FSs across NTN entities and ground stations (GSs), considering the constraints imposed by maximum separation distance. In this paper, our investigation focuses specifically on Low Earth Orbit (LEO) satellites, characterized by rapid orbital velocity and non-stationary positioning relative to ground users. 

The mobility of LEO satellites introduces challenges when deploying RAN across NTN, leading to frequent disruptions in establishing and terminating links between BB functions distributed across satellites and GSs. Therefore, this study also assesses the performance of Conditional Handover (CHO) procedures under various NTN scenarios, while assuming some feasible FS options. Performance metrics encompass the handover duration, disconnection time, and control traffic volume on each satellite link. This evaluation is crucial for understanding the advantages and drawbacks of specific FS.

This paper aims to provide a comprehensive overview of FSs in RAN, elucidating the challenges and possible solutions arising from integrating NTN to provide connectivity to ground users. By examining the feasibility of RAN FSs across LEO satellites and GSs, and evaluating the performance of CHO procedures, we offer insights into the evolving landscape of wireless communication technologies. The main contributions of this paper are as follows:
\begin{itemize}
    \item  Evaluation of various FS options within the RAN architecture, particularly in the context of a static NTN.
    \item  Detailed analysis of the constraints associated with FSs, including factors such as distance between network nodes, latency, BW, antennae, and processing capacity.
    \item  Performance evaluation of the CHO procedures in various mobile NTN scenarios with different FS deployments

\end{itemize}
\vspace{-2mm}

\vspace{-1mm}

\section{RAN Architectures}

\label{sec:RAN Arch}

\begin{figure}
     \centering
     \begin{subfigure}[t]{0.99\linewidth}
         \centering
         \includegraphics[width=\linewidth]{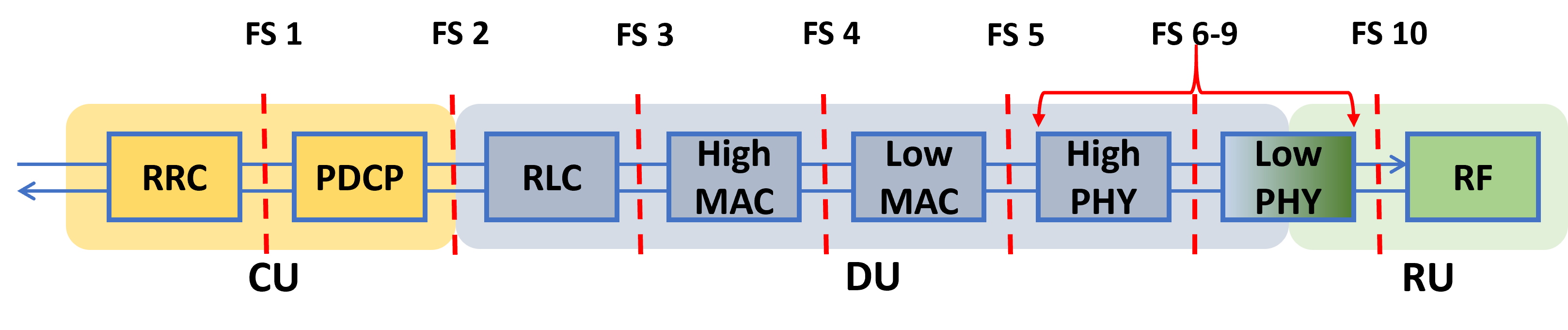}
         \caption{Baseband Function splits}
         \label{fig:1a}
     \end{subfigure}
     \hfill
     \begin{subfigure}[t]{0.99\linewidth}
         \centering
         \includegraphics[width=\linewidth]{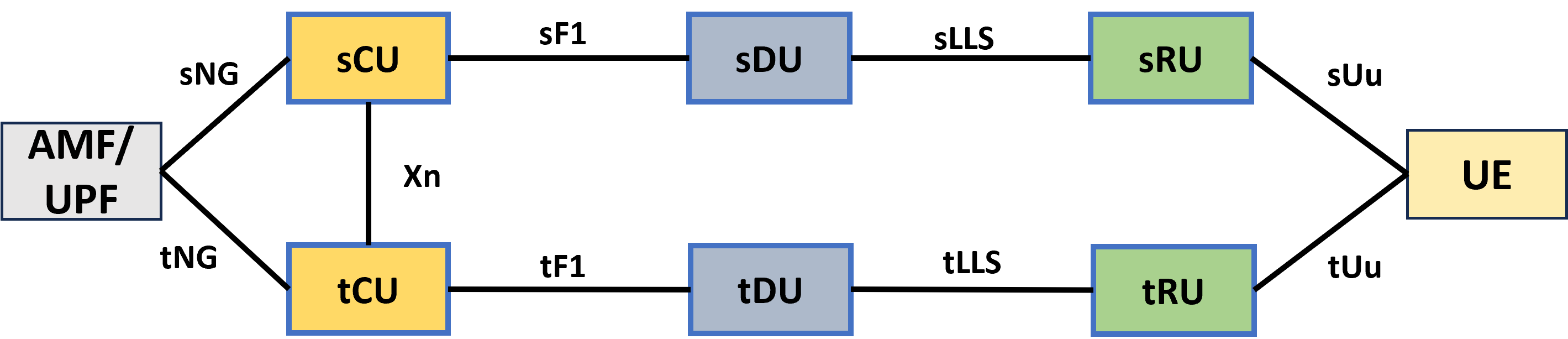}
         \caption{Reference architecture}
         \label{fig:1b}
     \end{subfigure}

    \vspace{-1mm}
    \caption{Overview of RAN architecture.}
    \label{fig:ran arch}
\end{figure}

The RAN architecture evolves with different generations of mobile communication technologies and forms an indispensable component of the mobile network architecture. However, the data packet or the request has to go through a stack of protocol layers where several functions like signal processing are performed and are known as BB functions. 
Three standardized entities are defined as a subset of BB functions namely Radio unit (RU), Distributed unit (DU), and Central unit (CU) as shown in Fig.~\ref{fig:1a}. According to Small Cell Forum (SCF)~\cite{r3}, $10$ FSs are depicted as shown in Fig.~\ref{fig:1a} where the red dotted line illustrates the FS option separating the BB functions among the entities. The functions to the right side of line are in the DU whereas the rest are in the CU.

The RU is responsible for Radio Frequency (RF) operations and signal processing and also hosts some lower physical layer functions depending on the FS option, \emph{i.e.,} FS $6$ to $9$. In contrast, functions to the right of FS $10$ solely represent RU without any physical layer (PHY) functions. In general, the DU consists of the functions Radio Link Control (RLC), Medium Access Control (MAC), and some functions from the physical layer depending on the FS option. However, the functions can be pulled or pushed to DU/CU depending on the FS, \emph{i.e.,} the DU may comprise Packet Data Convergence Protocol (PDCP) if FS $1$ is implemented and CU might comprise RLC if FS $3$ is implemented. The CU is generally responsible for the higher layer functions, namely PDCP and Radio Resource Control (RRC). From a downlink (DL) perspective, the PHY function in DU is Forward Error Correction (FEC) for FS $6$. In addition to FEC, the Quadrature Amplitude Modulation function (QAM) and antenna mapping are present in DU for FS $7$. Similarly, for FS $8$, resource mapping is added. For FS $9$, Inverse Fast Fourier Transform (IFFT), and cyclic prefix are placed on the DU additionally. 

The reference architecture utilized for our investigation in Section~\ref{sec:mobility and HO} is illustrated in Fig.~\ref{fig:1b}, with entities aligned according to color-coded references from Fig.~\ref{fig:1a}. The prefixes “s” and “t” are used to denote the ``source” and ``target” links or nodes. It can be seen that from Fig.~\ref{fig:1b}, the RU is connected to the User Equipment (UE) through the Uu interface or air interface. It receives instructions from the upper layers through a Lower-Layer Split (LLS) interface. A CU communicates with a DU using the F1 interface. Xn interface is specified for the communication between the two RANs. Furthermore, the Core Network (CN) is composed of many functions. However, we consider the Access and Mobility Function (AMF) and User Plane Function (UPF), as they are responsible for mobility and some control signaling and are connected to the RAN through the Next-generation (NG) interface as shown in Fig.~\ref{fig:1b}.

In a traditional base station, BB processing functions (CU+DU) are co-located with RU at the same site,  usually known as D-RAN architecture. This architecture reduces latency. However, this increases cost, in terms of power consumption and capital expenditure, in proportion to the cell sites~\cite{r6,r7, r8}. 
Subsequently, first-generation C-RAN emerged, segregating RUs at cell sites while centralizing BB processing in BB unit pools, connecting them with fronthaul links, however, each BB processing unit is designated with a specific RU. 
Next, BB processing is virtualized, called Virtualized RAN (vRAN) offering cost, performance, and deployment advantages due to shared BB processing, and reduced base station payload respectively~\cite{r1}. It minimizes hardware dependency by virtualizing BB functions and using general-purpose servers. Nonetheless, centralization of all BB processing functions presents challenges such as stringent fronthaul (FH) latency and heavy BW requirements in the transport network. The FH interface is traditionally called CPRI. However, in the later architectures, it is termed an LLS interface due to the lower BB FS between the entities.
Hybrid C-RAN (H-CRAN) in ~\cite{r4} uses the FS model based on cell and user processing functions to distribute the BB functions between edge (RU+DU) and central (CU) clouds. Despite these advancements, interfaces are proprietary between entities leading to vendor lock-in, except for the open F1 interface. Open-RAN (O-RAN) is a standardization effort to open the interfaces and evolve towards more intelligent vRAN ~\cite{r5}, allowing for interoperability between entities from different vendors, and introducing a RAN intelligent controller (RIC) for automated and optimized RAN operations.

\vspace{-2mm}
\section{Analysis of FSs in a Static NTN Scenario}
\vspace{-.5mm}
\label{sec:feasible splits}

In a static NTN scenario, which provides a momentary snapshot of an LEO satellite serving a single ground UE, connected to the GS through a Feeder Link (FL) at a specified elevation angle, our objective is to analyze the feasibility of different FS configurations depicted in Fig.~\ref{fig:1a}, linking the GS (CU) to the satellite (DU+RU). We assess various FS options on various dimensions such as FH latency, BW, and antennae. We analyze FS options, and navigate challenges posed by stringent latency requirements and increasing computational demands.

\begin{table*}[!ht]
\vspace{0.03in}
\centering
\caption{Function splits, Fronthaul parameters and Processing requirements.}
\label{tab:FH latency}
\begin{tabular}{|l|c|c|cccc|cc|}
\hline
\multicolumn{1}{|c|}{\multirow{3}{*}{\textbf{Function Split option}}} & \multirow{3}{*}{\textbf{\begin{tabular}[c]{@{}c@{}}One way FH latency \\ specifications (SCF~\cite{r3})\end{tabular}}} & \multirow{3}{*}{\textbf{\begin{tabular}[c]{@{}c@{}}Max. FH distance \\ in free space\end{tabular}}} & \multicolumn{4}{c|}{\textbf{\begin{tabular}[c]{@{}c@{}}Required FH bandwidth~\cite{r3}\\ {[}Mbps{]}\end{tabular}}} & \multicolumn{2}{c|}{\textbf{\begin{tabular}[c]{@{}c@{}}Processing\\ requirements~\cite{r4}\\ {[}GOPS{]}\end{tabular}}} \\ \cline{4-9} 
\multicolumn{1}{|c|}{} &  &  & \multicolumn{2}{c|}{\textbf{N$_{ant}$}=2} & \multicolumn{2}{c|}{\textbf{N$_{ant}$}=64} & \multicolumn{1}{c|}{\multirow{2}{*}{\textbf{GS}}} & \multirow{2}{*}{\textbf{Satellite}} \\ \cline{4-7}
\multicolumn{1}{|c|}{} &  &  & \multicolumn{1}{c|}{\textbf{DL}} & \multicolumn{1}{c|}{\textbf{UL}} & \multicolumn{1}{c|}{\textbf{DL}} & \textbf{UL} & \multicolumn{1}{c|}{} &  \\ \hline
1 – { RRC-PDCP } & Non-ideal- 30 ms & 9000 km & \multicolumn{1}{c|}{149.9} & \multicolumn{1}{c|}{48.6} & \multicolumn{1}{c|}{149.9} & 48.6 & \multicolumn{1}{c|}{\textless  8} & \textgreater{} 36.5 \\ \hline
2 – { PDCP - RLC } & Non-ideal- 30 ms & 9000 km & \multicolumn{1}{c|}{150} & \multicolumn{1}{c|}{48.7} & \multicolumn{1}{c|}{150} & 48.7 & \multicolumn{1}{c|}{\textless 8} & \textgreater{} 36.5 \\ \hline
3 – { RLC - MAC } & Non-ideal- 30 ms & 9000 km & \multicolumn{1}{c|}{150.6} & \multicolumn{1}{c|}{48.9} & \multicolumn{1}{c|}{150.6} & 48.9 & \multicolumn{1}{c|}{\textless 8} & \textgreater{} 36.5 \\ \hline
4 – { hMAC- lMAC } & Sub ideal- 6 ms & 1800 km & \multicolumn{1}{c|}{151.3} & \multicolumn{1}{c|}{49.4} & \multicolumn{1}{c|}{151.3} & 49.4 & \multicolumn{1}{c|}{\textless 8} & \textgreater{} 36.5 \\ \hline
5 – { MAC - PHY } & Sub ideal- 6 ms & 1800 km & \multicolumn{1}{c|}{152.3} & \multicolumn{1}{c|}{49.9} & \multicolumn{1}{c|}{152.3} & 49.9 & \multicolumn{1}{c|}{\textless 8} & \textgreater{} 36.5 \\ \hline
6 – { PHY Split I } & Near ideal- 2 ms & 600 km & \multicolumn{1}{c|}{173.1} & \multicolumn{1}{c|}{451.6} & \multicolumn{1}{c|}{173.1} & 451.6 & \multicolumn{1}{c|}{8} & 36.5 \\ \hline
7 – { PHY Split II  } & \begin{tabular}[c]{@{}c@{}}Near ideal- 2 ms \\ Ideal- 0.25 ms\end{tabular} & \begin{tabular}[c]{@{}c@{}}600 km\\ 75 km\end{tabular} & \multicolumn{1}{c|}{932.6} & \multicolumn{1}{c|}{903.2} & \multicolumn{1}{c|}{29843} & 28901 & \multicolumn{1}{c|}{15.9} & 28.6 \\ \hline
8 – { PHY Split III } & \begin{tabular}[c]{@{}c@{}}Near ideal- 2 ms\\ Ideal- 0.25 ms\end{tabular} & \begin{tabular}[c]{@{}c@{}}600 km\\ 75 km\end{tabular} & \multicolumn{1}{c|}{1075.2} & \multicolumn{1}{c|}{921.6} & \multicolumn{1}{c|}{34406} & 29491 & \multicolumn{1}{c|}{18.5} & 26 \\ \hline
9 – { PHY Split IIIb } & \begin{tabular}[c]{@{}c@{}}Near ideal- 2 ms\\ Ideal- 0.25 ms\end{tabular} & \begin{tabular}[c]{@{}c@{}}600 km\\ 75 km\end{tabular} & \multicolumn{1}{c|}{1966.1} & \multicolumn{1}{c|}{1966.1} & \multicolumn{1}{c|}{62915} & 62915 & \multicolumn{1}{c|}{19.8} & 24.7 \\ \hline
10 – { PHY Split IV/PHY - RF } & Ideal- 0.25 ms & 75 km & \multicolumn{1}{c|}{2457.6} & \multicolumn{1}{c|}{2457.6} & \multicolumn{1}{c|}{78643} & 78643 & \multicolumn{1}{c|}{23.8} & 20.7 \\ \hline
\end{tabular}
\vspace{-6mm}
\end{table*}

Table~\ref{tab:FH latency} illustrates the $10$ FS options alongside their respective FH latency requirements~\cite{r3} for different use cases. Full gNB onboard the satellite is not tabulated since there will be only backhaul and no FH in this deployment. Furthermore, the backhaul constraints are not that stringent when compared to the FH. The maximum FH distance is determined as the product of the speed of light in free space and the FH latency requirement. The separation distance is defined as the distance between the satellite and the GS. The FH BW is computed utilizing the equations and parameters outlined in~\cite{r3}. Additionally, the processing requirements, measured in Giga operations per second (GOPS), are computed based on the specifications provided in~\cite{r4}, which is the summation of user processing functions and cell processing functions at GS and satellite after the split. The user processing functions include higher layer functions namely PDCP, RLC, MAC, and PHY functions namely FEC, QAM, and antenna mapping. The cell processing functions include remaining low PHY functions as mentioned in Section~\ref{sec:RAN Arch}. Note that these processing requirements are calculated under the assumption of a single user, thus the cell processing requirement is not significant enough. In scenarios with multiple users, the allocated GOPS for cell processing remains constant for all users, although the user processing requirement increases with the users.

In Table~\ref{tab:FH latency}, as the FS shifts to lower layers, the FH latency becomes more stringent. An FS is said to be feasible if the separation distance is less than the maximum FH distance. Therefore, a LEO satellite orbiting at an altitude of $600$ km, with FL elevation angles of $10^\circ$ and $90^\circ$, will exhibit separation distances approximately measuring $1935$ km and $600$ km, respectively (as depicted in Table~\ref{tab:prop delay}). Hence, the feasible FSs in the former case are $1$-$3$ whereas in the latter case, the feasible FSs are $1$-$9$, assuming near ideal use case for the PHY layer splits. A large FH distance indicates a relaxed latency requirement which may translate to, a single hop with said distance or multiple hops as long as the total separation distance is less than the FH distance.

From Table~\ref{tab:FH latency}, with a varying number of antenna (N$_{ant}$)~\cite{r3}, the FH BW requirement in both DL and uplink (UL) increases as the FS moves to lower layers. The FH BW requirement for N$_{ant}=2$, almost increased by $5.4$ times in DL and doubled in UL when the FS changed from $6$ to $7$. Similarly, for N$_{ant}=64$, the FH BW increased by $173$ times in DL and in UL by about $64$ times when the FS changed from $6$ to $7$. Furthermore, the BW requirements in DL and UL are about the same for the FSs $1$ to $5$ for both cases as the BW requirement in higher layers is independent of N$_{ant}$~\cite{r3}. Table~\ref{tab:FH latency} shows that FS $10$ (PHY-RF) necessitates approximately a $2.4$ Gbps and $78.6$ Gbps FH link rate respectively, for both DL and UL within one beam of the satellite, serving a single user. Hence, for multiple beams of a single satellite, the required BW increases and might be a bottleneck for the feasibility of an FS. Furthermore, Table~\ref{tab:FH latency} shows that the computational requirements onboard the satellite decrease as the FS moves to lower layers. Similarly, the computational requirements at the GS increase as the FS moves to lower layers and vice versa.

The discussion above highlights how separation distances limit the feasibility of certain FS options in NTN, especially in LLS (FS $7$-$10$). While FH latency specifications are based on LTE with advanced TN features, they may not directly apply to legacy New Radio (NR) NTN. However, these specifications can be adapted to NR NTN by adjusting certain features. In legacy NR, FH latency requirements may be relaxed using asynchronous Hybrid Automatic Repeat Request (HARQ) specifications instead of LTE synchronous HARQ. Moreover, in NTN Rel-17~\cite{3GPP_NTN}, completely disabling HARQ feedback is permitted due to significantly longer round trip delays.

Furthermore, one of the major bottlenecks in the FH latency specifications from SCF is channel state information (CSI) reporting or reciprocity-based channel estimation, which requires a very short delay. It is unlikely that these kinds of features will be used in legacy NR NTN as they only boost data rates in good terrestrial coverage or in time division duplex (TDD) deployments. If disabled, they do not prevent basic functionalities. Hence, we can come to a consensus that these kinds of features are not necessary in NTN.

\begin{figure*}[h]
    \centering
    \includegraphics[width={0.9\linewidth}]{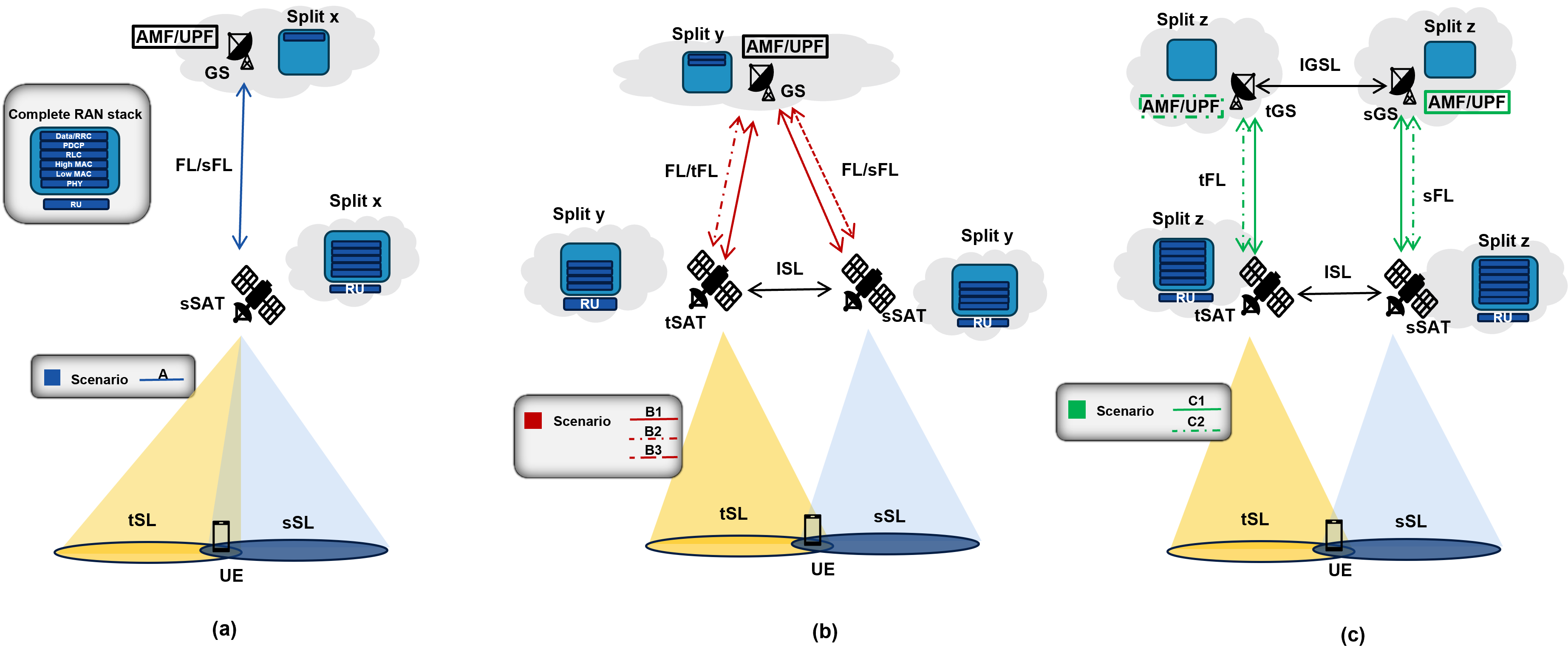}
    \vspace{-2.5mm}
    \caption{Reference scenarios (a) A - 1 satellite and 1 GS, (b) B - 2 satellites and 1 GS, (c) C - 2 satellites and 2 GSs.}
    \label{fig:model}
    \vspace{-6mm}
\end{figure*}

\vspace{-.5mm}
\section{Analysis of FSs based on mobile NTN scenarios}
\label{sec:mobility and HO}
\vspace{-.5mm}

In this section, we explore three distinct network scenarios (labeled A, B, and C), each showcasing unique configurations of satellite, ground station, and FS deployment.

Our scenario comprises two cells: a source cell (blue in Fig.~\ref{fig:model}) where a user originates and a target cell (orange in Fig.~\ref{fig:model}) where the user moves. It includes at least one satellite, one GS connected to the core network, and various satellite links: Service Link (SL) between a satellite and the user, FL between a satellite and its serving GSs, Inter-Satellite Link (ISL) connecting two satellites, and Inter-GS Link (IGSL) connecting two GS. The prefixes ``s" and ``t" are used to denote the ``source" and ``target" links in which the traffic flows before or after the user performs the CHO procedure, respectively.

In this study, three scenarios depicted in Fig.~\ref{fig:model} are evaluated, each representing a specific network topology, and for each scenario, three FS deployments, which are LLS, CU-DU split and gNodeB (gNB) on-board are analyzed. LLS option which may be referred to FS $8$ in Fig.~\ref{fig:1a} since there is no clear consensus regarding LLS. As discussed in Section~\ref{sec:feasible splits}, LLS is infeasible FS for ideal use cases. However, we assume that certain features are relaxed, rendering this FS feasible. Secondly, the CU-DU split can be referred to FS $2$ in Fig.~\ref{fig:1a}. Thirdly, the gNB on-board option is not an actual FS represented in Fig.~\ref{fig:1a}, where all functions apart from AMF/UPF are on the satellite as in Fig.~\ref{fig:1b}. To interpret the model for any sub-scenario in Fig.~\ref{fig:model}, begin from the GS, indicated by a specific legend (\emph{e.g.}, B2), which connects to the target satellite and is linked to the source satellite via ISL, while disregarding other legends. The depiction of the FS between the GS and the satellite is shown in Fig.~\ref{fig:model}, where split x/y/z can represent one of the three split options mentioned previously.

In Fig.~\ref{fig:model}(a), scenario A is characterized by a single satellite serving both cells. Given a constellation comprising several hundreds of satellites, each tasked with serving a substantial number of cells, this scenario is the most probable. In scenario B in Fig.~\ref{fig:model}(b), the cells are handled by different satellites, however, both are connected to the same GS in different ways. This is typical at the edge of two areas handled by the two satellites. For scenario C in Fig.~\ref{fig:model}(c), there are also two distinct GSs; however, a single AMF is still co-located with one of the two GSs. This case is less likely than the other ones yet it may happen at the edge of two areas handled by the same GS, \emph{e.g.}, at the border of two countries. In scenario C, it should be noted that there are always two full gNBs regardless of the FS as their highest layers are either on the two satellites or on the two GSs, but never in a common network node.

Assuming a specific altitude and elevation angle, propagation delays for the SL and FL can be estimated using trigonometric principles. Similarly, if a certain number of satellites orbiting in the same plane then the distance between the satellites can be calculated, thus determining the ISL delay. Finally, the two GSs will lie on the edge of the satellite footprints as defined by the FL elevation angle.  
On average, we can infer that they are separated by the same angle, relative to the Earth's center, as the two satellites, facilitating the computation of the geodesic distance. To accommodate factors such as non-vacuum propagation and indirect paths caused by router usage, we can augment this distance by $20\%$ before dividing it by the speed of light to calculate the propagation delay. Table~\ref{tab:prop delay} shows the propagation delay of each satellite link considering various elevation angles at $600$ km of altitude where the low elevation angles represent worst-case scenarios. The highlighted row indicates the values utilized in the subsequent part of the paper.

\begin{table}[t!]
\centering
\caption{Different link propagation delays.}
\vspace{-2mm}
\label{tab:prop delay}
\begin{tabular}{|c|c|c|c|c|c|}
\hline
{\color[HTML]{000000} \textbf{Constellation}} & {\color[HTML]{000000} \textbf{\begin{tabular}[c]{@{}c@{}}Elevation\\ Angles\end{tabular}}} & {\color[HTML]{000000} \textbf{\begin{tabular}[c]{@{}c@{}}SL\\ delay\\ {[}ms{]}\end{tabular}}} & {\color[HTML]{000000} \textbf{\begin{tabular}[c]{@{}c@{}}FL\\ delay\\ {[}ms{]}\end{tabular}}} & {\color[HTML]{000000} \textbf{\begin{tabular}[c]{@{}c@{}}ISL\\ delay\\ {[}ms{]}\end{tabular}}} & {\color[HTML]{000000} \textbf{\begin{tabular}[c]{@{}c@{}}IGSL\\ delay\\ {[}ms{]}\end{tabular}}} \\ \hline
\rowcolor[HTML]{FBE4D5} 
{\color[HTML]{000000} \textbf{LEO@600}} & {\color[HTML]{000000} \begin{tabular}[c]{@{}c@{}}SL: 30 deg\\ FL: 10 deg\end{tabular}} & {\color[HTML]{000000} 3.59} & {\color[HTML]{000000} 6.45} & {\color[HTML]{000000} 7.28} & {\color[HTML]{000000} 7.99} \\ \hline
{\color[HTML]{000000} \textbf{LEO@600}} & {\color[HTML]{000000} \begin{tabular}[c]{@{}c@{}}SL: 90 deg\\ FL: 90 deg\end{tabular}} & {\color[HTML]{000000} 2.00} & {\color[HTML]{000000} 2.00} & {\color[HTML]{000000} 7.28} & {\color[HTML]{000000} 7.99} \\ \hline

\end{tabular}
\end{table}

\begin{figure*}
     \centering
     \begin{subfigure}[b]{0.28\textwidth}
         \centering
         \includegraphics[width=\textwidth]{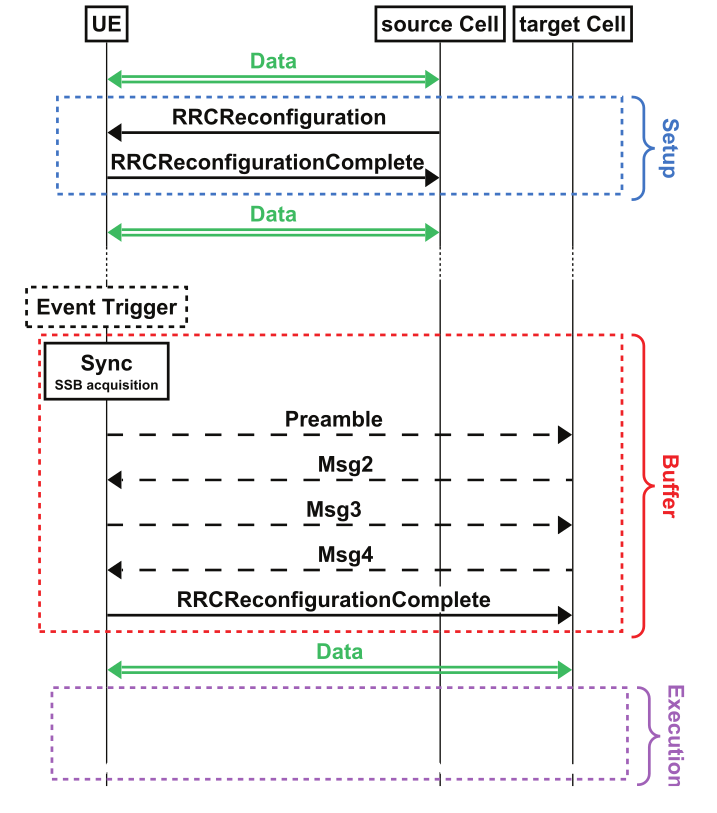}
         \caption{Intra-DU CHO procedure}
         \label{fig:Intra DU}
     \end{subfigure}
     \hfill
     \begin{subfigure}[b]{0.32\textwidth}
         \centering
         \includegraphics[width=\textwidth]{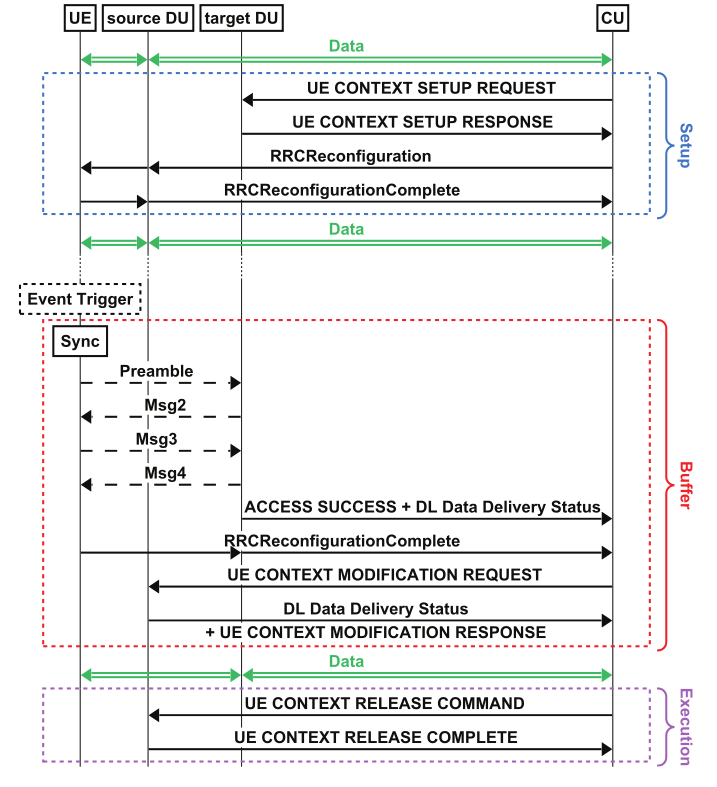}
         \caption{Inter-DU CHO procedure}
         \label{fig:Inter DU}
     \end{subfigure}
     \hfill
     \begin{subfigure}[b]{0.38\textwidth}
         \centering
         \includegraphics[width=\textwidth]{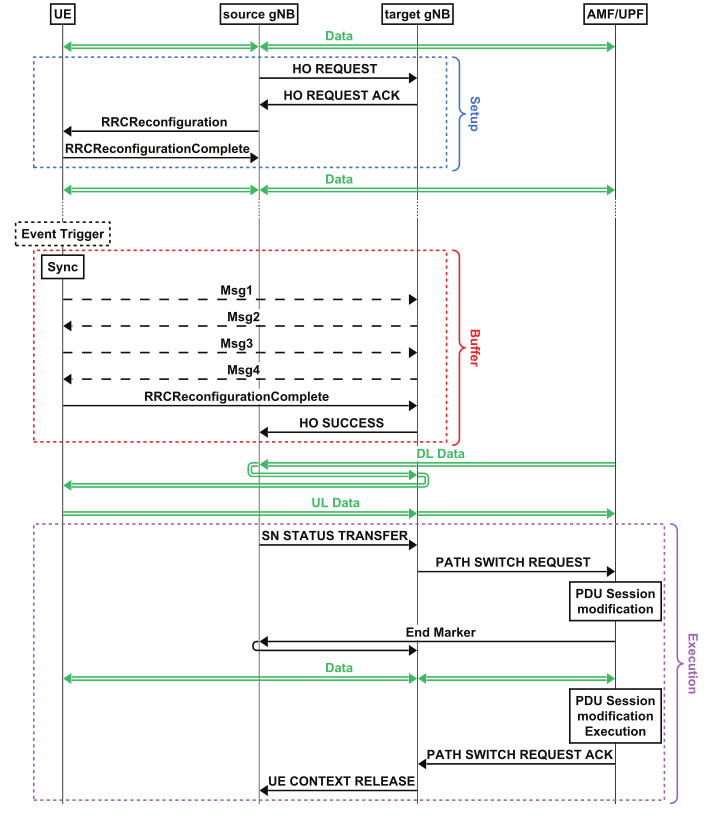}
         \caption{Inter-gNb Intra-AMF CHO procedure}
         \label{fig:Xn}
     \end{subfigure}
    \vspace{-1.5mm}
    \caption{Complete message exchange associated with each procedure.}
    \label{fig:Message exchange}
    \vspace{-6mm}
\end{figure*}

For the system architecture, we consider two chains of gNB functions, one serving the source cell and one serving the target cell, the AMF/UPF as representative of the CN as shown in Fig.~\ref{fig:1b}. Depending on the scenario, corresponding functions in the source and target chain may also be merged, in which case the UE performs a specific variant of the mobility procedure within the same gNB unit. For instance, in the case of the LLS option where two satellites and only one GS is involved, the two RUs on the satellites (sRU and tRU) may be connected to a common DU deployed in the GS. The UE would perform an Intra-DU CHO procedure.

All procedures have the same behavior towards the UE, \emph{i.e.,} the signaling over the tUu and sUu interfaces is the same in all cases, which consists of the RRC CHO procedure as seen in Fig. \ref{fig:Message exchange}. It can be divided into three phases: setup, buffer, and execution.
During the setup phase, the sgNB configures the UE with a specific trigger condition related to a certain metric, \emph{e.g.}, time, distance, or target and source cell signal strength. 
The UE then starts to perform measurements regarding the configured metric. When the condition is triggered, the buffer phase can begin.
During the buffer phase, the UE synchronizes with the tgNB by acquiring its Synchronization Signal Block (SSB) and performs the Random Access Procedure. The UE cannot deliver data as the RRC connection is re-established only after Msg4 is delivered. Hence, this phase duration is a key metric to assess the quality of the FSs. The UE resumes data delivery, often requiring additional signaling in the execution phase to ensure correct network configuration. This phase typically involves signaling through gNB internal or NG interface~\cite{3GPP_RRC}.

When the DU or RU is common to both source and target chains, the Intra-DU CHO procedure is performed which consists only of the already mentioned RRC CHO procedure over Uu~\cite{3GPP_RRC}. When the CU is common to both source and target chains, the Inter-DU CHO procedure is performed. During the setup phase, before configuring the UE, the CU requests all the potential tDUs if they can accept the UE through the F1 interface. After all replies are collected, the CU can configure the UE through the source chain. In the buffer phase after the UE has completed the RACH procedure, the tDU informs the CU of the successful access so that the CU can inform the sDU that it can stop data delivery. Finally, in the execution phase, the CU informs the sDU that it can release the UE context~\cite{3GPP_ARCH}. Furthermore, the Inter-gNB Intra-AMF CHO is performed as the mobility happens between two complete gNBs. This procedure may be performed over the Xn or NG interface depending on the deployment. Here, we assume it will be performed over Xn for the sake of simplicity. Like the previous procedure, the sgNB asks the potential tgNBs if they can accept the UE although now this happens through the Xn interface. During the buffer phase, after the UE has triggered the procedure and has completed the RACH, the tgNB informs the sgNB of the successful access. At this point, the UE can resume data delivery, however, the network User Plane Function (UPF) is not informed yet that the serving gNB has changed. Hence, it will keep forwarding DL data to the sgNB which momentarily will forward it further to the tgNB which will deliver it to the UE, causing an additional delay. The UL data is not affected as the tgNB will still forward it directly to the UPF. To fix this, during the execution phase, the tgNB triggers a Path Switch request which involves several Application Programming Interface (API) calls to various CN functions and the UPF followed by the propagation of an End Marker. When the tgNB receives the confirmation of a successful path switch, it can inform the sgNB that it can release the UE context. It is also assumed that all network nodes, source, and target cells belong to the same Tracking Area (TA) so that a TA Update procedure is not performed.

Fig.~\ref{fig:Message exchange} shows the complete message exchange associated with each procedure. Depending on the mapping between each architecture interface and satellite interface, which depends on the FS and scenario considered, it is possible to determine the propagation delay associated with each message transmission. Furthermore, it is assumed that the reception of each message requires a $1$ ms processing delay, the SSB acquisition requires $20$ ms, and the CN API calls performed in the Inter-gNB Intra-AMF CHO procedure require a total of $50$ ms.

\vspace{-1mm}
\section{Results and analysis}
\label{sec: Results}
\begin{figure}[t]
    \centering
    \includegraphics[width={0.99\linewidth}]{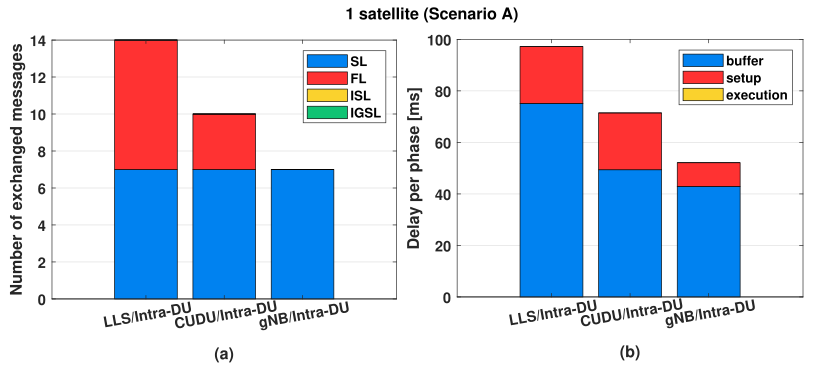} 
    \caption{CHO performance for scenario A.}
    \label{fig:HO1}
    \vspace{-5mm}
\end{figure}

For Scenario A, regardless of the FS, the Intra-DU CHO procedure is always performed as all functions in the two chains are in common. From Fig.~\ref{fig:HO1}(a), it is evident that relocating RAN functions onboard the satellite results in fewer messages directed to the GS, thereby eliminating the need for traversal through both the SL and the FL for delivery. This substantially reduces the overall procedure duration, particularly the time spent in a buffer state as shown in Fig.~\ref{fig:HO1}(b).

\begin{figure}[t!]
    \centering
    \includegraphics[width={0.99\linewidth}]{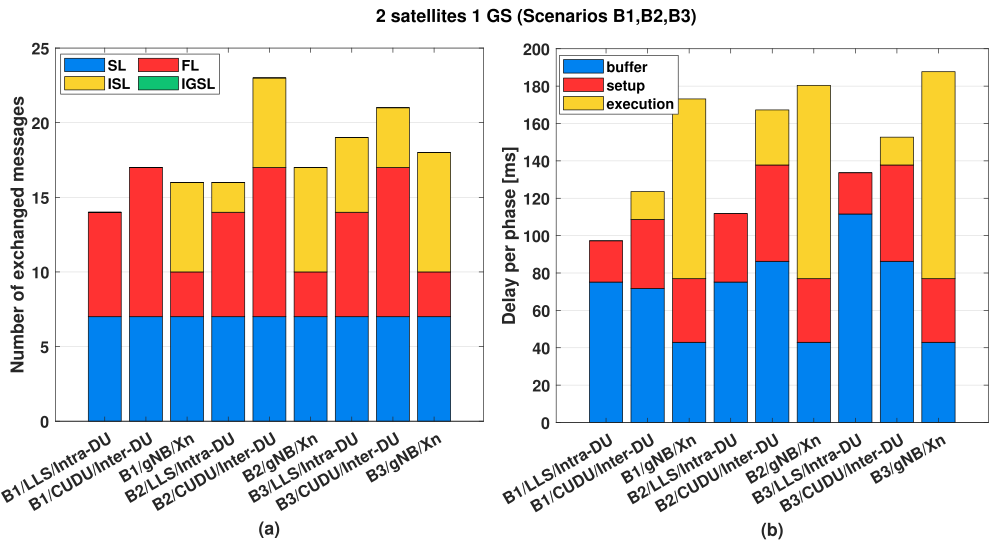}
    \caption{CHO performance for scenario B.}
    \label{fig:HO2}
    \vspace{-5mm}
\end{figure}

In scenario B, each bar in Fig.~\ref{fig:HO2} corresponds to a distinct combination of scenario variant, FS, and CHO procedure variant. The trend is opposite as compared to Fig.~\ref{fig:HO1}, as the more complexity is moved onto the satellite, the longer the procedure takes to complete as shown in Fig.~\ref{fig:HO2}(b). This is mainly due to the higher complexity of the procedure itself, the Inter-gNB Intra-AMF CHO also involves a considerable amount of time spent in performing the Path Switch procedure, unlike the other procedures. Each split option entails a unique RAN function common at the GS, resulting in a distinct procedure. However, it is notable that the time spent in the buffer state exhibits an opposite trend, being the shortest with the gNB onboard satellite option, due to similar reasons as observed in scenario A. Additionally, the CU-DU split results in the highest volume of messages sent, both in FL and in overall links as shown in Fig.~\ref{fig:HO2}(a). FL is typically the network bottleneck due to the convergence of traffic from numerous cells onto the same link, potentially posing capacity challenges for mobility procedures with the CU-DU split.


\begin{figure}[t!]
    \centering
    \includegraphics[width={0.99\linewidth}]{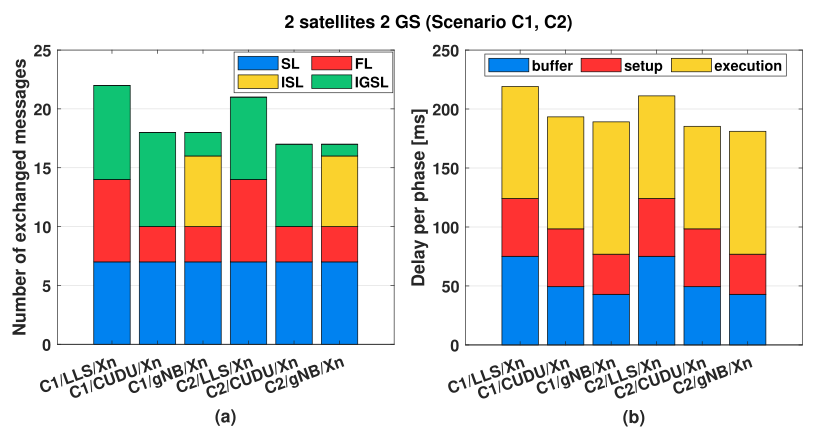}
    \caption{CHO performance for scenario C.}
    \label{fig:HO3}
    \vspace{-1mm}
\end{figure}

In scenario C, from Fig.~\ref{fig:HO3}(a), it is observed that the LLS option exhibits the highest volume of traffic transmitted over FL. Nevertheless, in this case, the capacity is not significantly impacted, as the traffic from LLS is largely independent of user/control traffic but rather depends more on the number of cells managed by the satellite(s), as given in Table~\ref{tab:FH latency}. Furthermore, from Fig.~\ref{fig:HO3}(b), it is evident that as more complexity is moved onboard the satellite, both the procedure duration and the time spent in the buffer state decrease.

The gNB on-board split option appears to be the most effective in reducing the overhead traffic across the network and minimizing the duration of the procedure, particularly in decreasing the time spent in the buffer state, where the data delivery pauses. The reduction of control overhead traffic is significant in NTN, where the mobility events predominantly occur due to changes in serving satellites for specific cells or due to the cell movement in the case of Earth-moving cells. These events typically occur every few minutes, prompting all the users within the cell simultaneously to execute mobility procedures. While strategies such as aggregating the control traffic or distributing the handovers over time can help mitigate this issue, a significant amount of traffic is still expected.
\section{Conclusions}
\label{sec: Conclusions}

In this paper, we analyzed the feasible FSs for a static NTN scenario focusing on factors such as FH latency, separation distance, elevation angle, FH BW and antennae. Our findings indicate that the stringent FH latency poses limitations on the feasibility of FSs, whereas the elevation angle of the FL exerts a more significant impact on this feasibility. We also explored some solutions where certain features that could be relaxed or disabled such as HARQ and CSI reporting to adapt LTE specifications into the legacy NR NTN, thus making LLSs feasible. Our analysis further highlights that LLS are subject to strict separation distance constraints and experience prolonged delays during mobility CHO procedures, notably affecting user data transmission interruptions and control traffic volumes across the FL. Moreover, the LLS option exhibits minimal computational demands onboard satellites, promising a reduction in power consumption and payload. Conversely, higher-layer splits, particularly those involving full gNB onboard satellites, exhibit relaxed bandwidth and delay constraints, enabling coverage of more distant areas from serving GSs at the expense of increased CPU and payload complexity.

\section*{Acknowledgement}
\vspace{-.5mm}
This work was supported in part by the CELTIC-NEXT Project, 6G for Connected Sky (6G-SKY), with funding received from the Vinnova, Swedish Innovation Agency.

\end{document}